\documentclass[final,1p,times]{elsarticle} 
\usepackage{graphicx} 
\usepackage{amssymb} 
\usepackage{amsthm} 
\usepackage{lineno} 

\journal{Nuclear Physics A} 
\begin{document} 

\begin{frontmatter} 


\title{Results on angular distributions of thermal dileptons in nuclear collisions}

\author{Gianluca Usai$^{a}$ for the NA60 collaboration}

\address[a]{Universit\`a and INFN Cagliari, 
Complesso Universitario di Monserrato,
09042 Monserrato (CA), Italy}

\begin{abstract} 
The NA60 experiment at the CERN SPS has studied dimuon production in
158 AGeV In-In collisions. The strong pair excess above the known
sources found in the mass region $0.2<M<2.5$ GeV has been previously
interpreted as thermal radiation. In this paper results on the
associated angular distributions for $M<1$ GeV, as measured in the
Collins-Soper reference frame, are presented. The structure function
parameters $\lambda$, $\mu$, $\nu$ are consistent with zero and the
projected polar and azimuth angle distributions are uniform. The
absence of any polarization is consistent with the interpretation of
the excess dimuons as thermal radiation from a randomized system.

\end{abstract} 

\end{frontmatter} 



Lepton pairs are a particularly attractive observable in high energy
nuclear collisions, because their continuous emission probes the
entire space-time evolution of the produced fireball.  To the extent
that the bulk consituents of the expanding matter (hadrons and
partons) equilibrate, the directly generated lepton pairs should
appear as a ``thermal radiation''.  Such a thermal radiation should
exhibit a number of features: (i) a Planck-like exponential shape of
mass spectra (strictly correct in case of a flat spectral function),
(ii) exponential $m_T$ spectra, (iii) the absence of any polarization
in the angular distributions.

Previous work on the mass spectra showed that dilepton production for
$M<1$ GeV is largely mediated by the process $\pi\pi\to \rho\to
\mu\mu$ with a strongly broadened $\rho$~\cite{NA60-1}. The analysis
of the $p_T$ spectra shows an increase of $T_{eff}$ vs $M$ up to $\sim
1$ GeV, while the drop at $\sim 1$ GeV and above might signal the
transition to the partonic thermal process $q\bar q\to\gamma\to
\mu\mu$~\cite{NA60-2, NA60-3}. The acceptance-corrected mass spectrum
falls exponentially and it is described quantitatively by theoretical
models which are explicitely based on medium thermalization up to 2.5
GeV~\cite{NA60-3}.

This paper concentrates on the first measurement of dilepton angular
distributions in high energy nuclear collisions. Results were already
recently published in~\cite{NA60-4}. Here, further details and aspects
on the results will be provided.  In general, the differential decay
angular distribution in the rest frame of the virtual photon with
respect to a certain set of axes can be written as~\cite{Gottfried}

\begin{equation}
\frac{1}{\sigma} \frac{d^2\sigma}{d\cos\theta d\phi} = 
\left(1 + \lambda\cos^2 \theta  + \mu \sin2\theta\cos\phi + \frac{\nu}{2}\sin^2\theta\cos2\phi 
\right).
\end{equation}

Here $\lambda$, $\mu$ and $\nu$ are structure functions related to
helicity structure functions and the spin density matrix elements of
the virtual photon. The nomenclature follows~\cite{Falciano,
Brandenburg}. Because of lack of sufficient statistics for $M>1$ GeV,
the present analysis is restricted to $M<1$ GeV, which is dominated by
pion annihilation.  Even in that case, where the annihilating
particles are spinless, the structure functions can be different from
zero because of orbital momentum. However, a completely random
orientation of annihilating pions in 3 dimensions, as expected in a
thermalized medium, should lead to $\lambda,$ $\mu$, $\nu=0$.

The decay angular distributions were studied in the Collins-Soper (CS)
reference frame~\cite{Collins-Soper}. The z axis is the bisector
between the beam and the negative target momenta (defining
the reaction plane). The polar angle $\theta$ is the angle between the
positive muon and the z-axis (defining the decay plane), while $\phi$
is the angle between the reaction and decay planes.  The choice of the
frame is not relevant: once all measured, $\lambda$, $\mu$ and $\nu$
can be re-computed in any other frame with a simple
transformation~\cite{Falciano}. This point will be illustrated in the
results.

Details of the NA60 apparatus are contained in~\cite{NA60-3,NA60-5}.
The data sample for 158 AGeV collisions is the same as
in~\cite{NA60-1, NA60-2}.  The analysis was done in the 2-dimensional
$\cos\theta-\phi$ space with different binnings in $(d^2N/d\cos\theta
d\phi)_{ij}$ in order to assess the systematics. For each {$i$,$j$} bin, the
analysis proceeded with the following steps: (i) subtraction of the
combinatorial background from $\pi$ and $K$ decays by the event-mixing
technique, (ii) subtraction of fake matches (incorrect associations of
muon tracks to the tracks reconstructed in the silicon pixel vertex
telescope), (iii) isolation of the excess by subtraction of the known
sources, (iv) acceptance correction in the 2-dimensional bin. Further
details for the analysis steps can be found in~\cite{NA60-1, NA60-2,
NA60-3, NA60-4}.  In order to increase the statistical significance,
no centrality selection was required. In order to exclude the region
of the low $m_T$ rise seen at all masses ~\cite{NA60-2}, a cut
$p_T>0.6$ GeV was applied. That has also the effect of improving the
background/signal ratio, which is $\sim$2-3.  The excess dileptons
were studied in two mass windows: $0.4<M<0.6$ GeV ($\sim17600$
$\mu\mu$ pairs) and $0.6<M<0.9$ GeV ($\sim36000$ $\mu\mu$ pairs). 
A fully differential acceptance correction should be performed in the
5-dimensional space $M-p_T-y-\cos\theta-\phi$. In this analysis the
Monte Carlo was first tuned with an iterative procedure to the measured $M$,
$y$ and $p_T$ spectra and then the acceptance matrix in the
$\cos\theta-\phi$ space was determined. The Monte Carlo muons were
overlayed on real data to include the effects of pair reconstruction
efficiencies.

\begin{figure}[ht]
\centering
\includegraphics[scale=0.32]{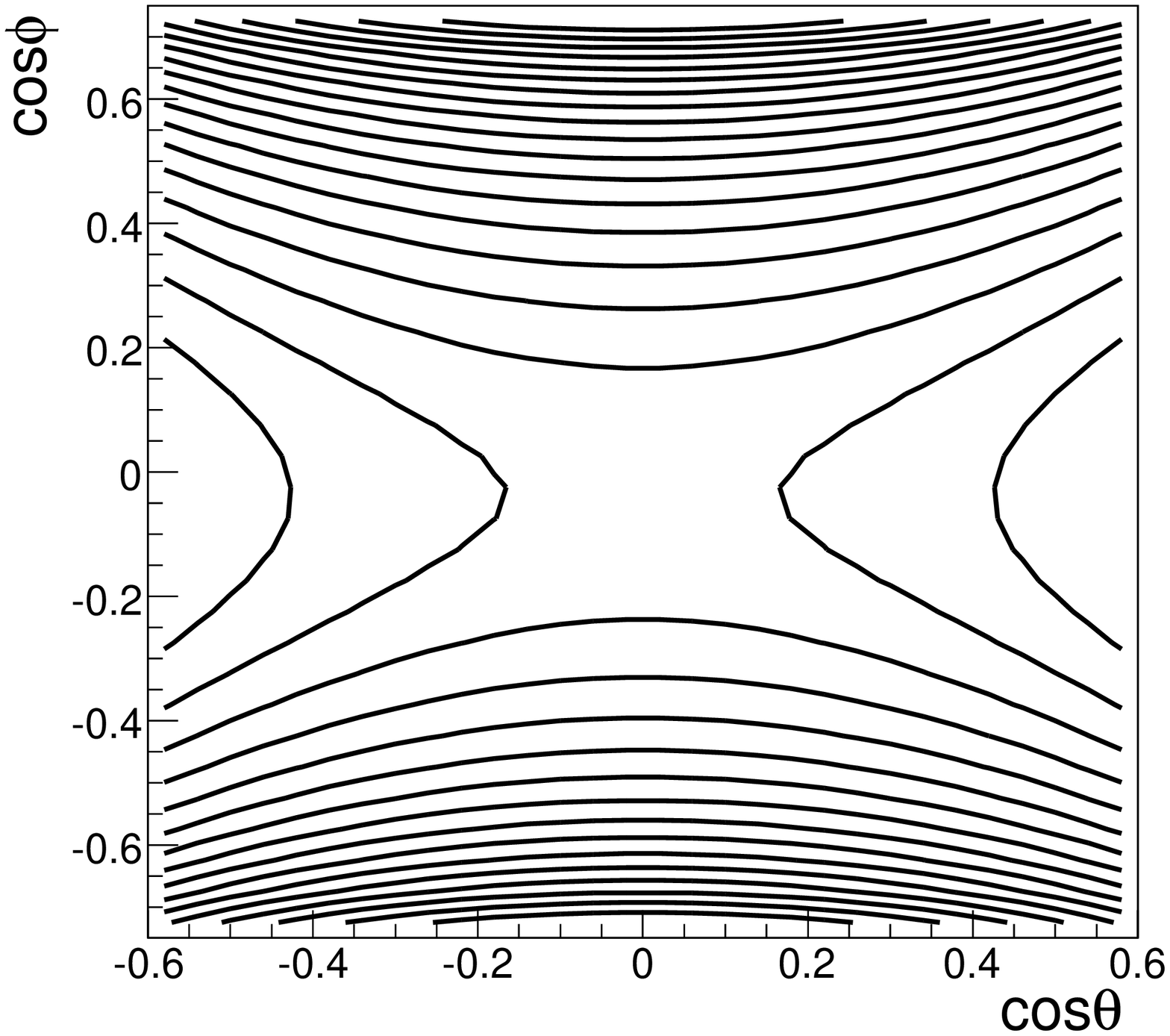}	 
\includegraphics[scale=0.32]{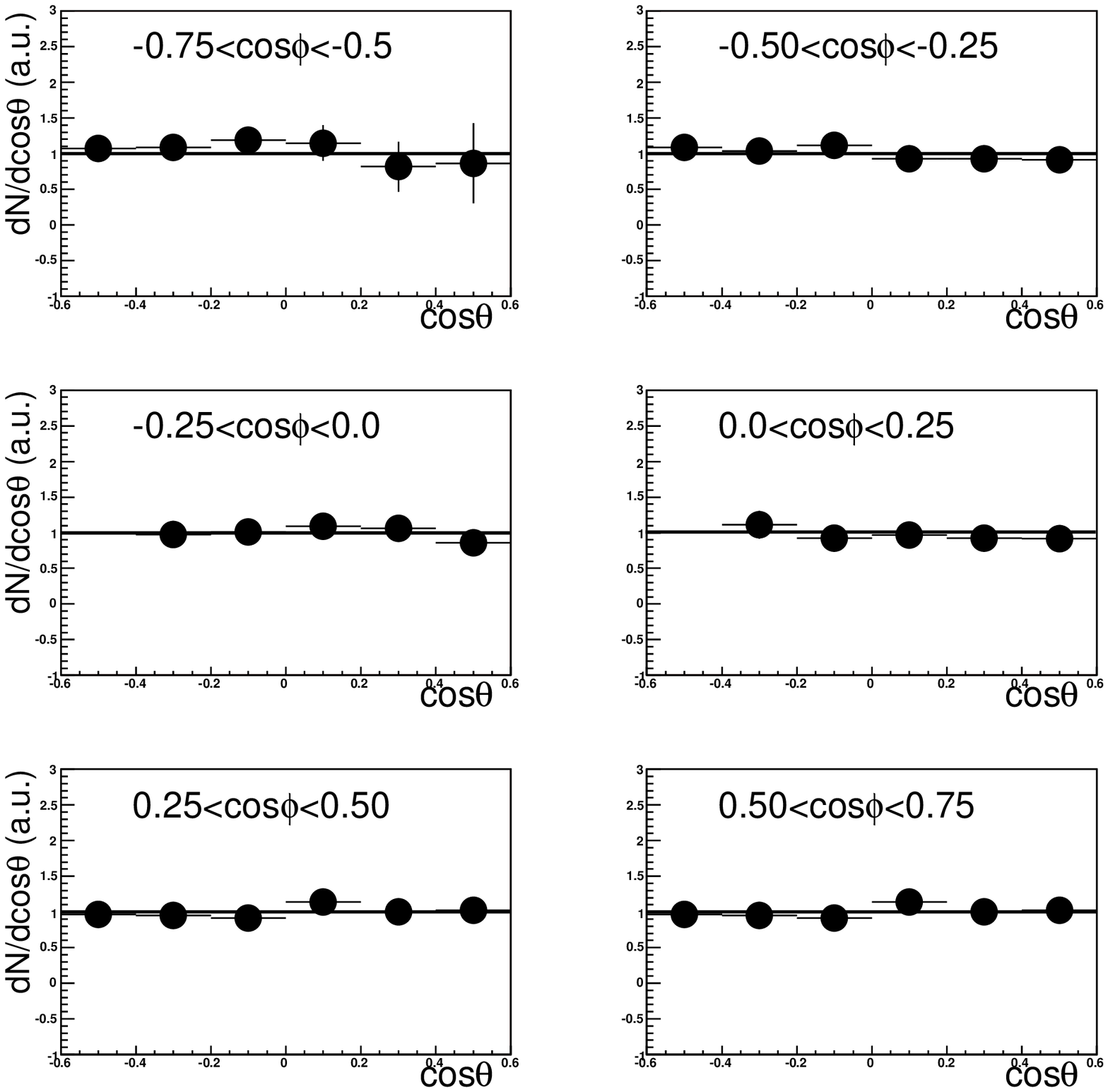}
\caption[]{Left: contour plot of bidimensional fit of $d^2N/d\cos\theta d\phi$ for $0.6<M<0.9$ GeV.
Right: Projections in $\cos\theta$ for the different bins in
$\phi$ for $0.6<M<0.9$ GeV.}
\label{Fig1}
\end{figure}

Three different methods were applied to extract the structure
functions.  The 2-dimensional distribution $d^2N/d\cos\theta d\phi$ was
directly fitted with Eq. (1). In order to avoid very low acceptance
bins, the 2-dimensional distribution was restricted to a 6x6 matrix
with $-0.6<\cos\theta<0.6$ and $-0 .75<cos\phi <0.75$.  Alternatively,
fixing $\mu$ to the value found in the 2-dimensional fit, the
2-dimensional distribution can be projected in $\cos\theta$ and $\phi$
and these 1-dimensional projections can be fitted with $ dN/d|\cos
\theta| \propto \left(1 + \lambda \cos ^2 \theta \right) $ and
$dN/d|\phi | \propto (1 + \lambda/3 +$$ \nu/3\cos 2\phi) $,
respectively (method 2). Finally, an analysis of the inclusive
distributions in $|\cos\theta|$ and $|\phi| $ can be also performed
(method 3).  In the analysis of 1-dimensional projections (methods 2
and 3), the range in $\theta$ and $\phi$ can be enlarged to
$|\cos\theta|<0.8$, $0<|\cos\phi|<0.75$ for method 2 and
$|\cos\theta|<0.8$, $0<|\phi|<\pi$ for method 3.


\begin{figure}[ht]
\centering
\includegraphics[scale=0.32]{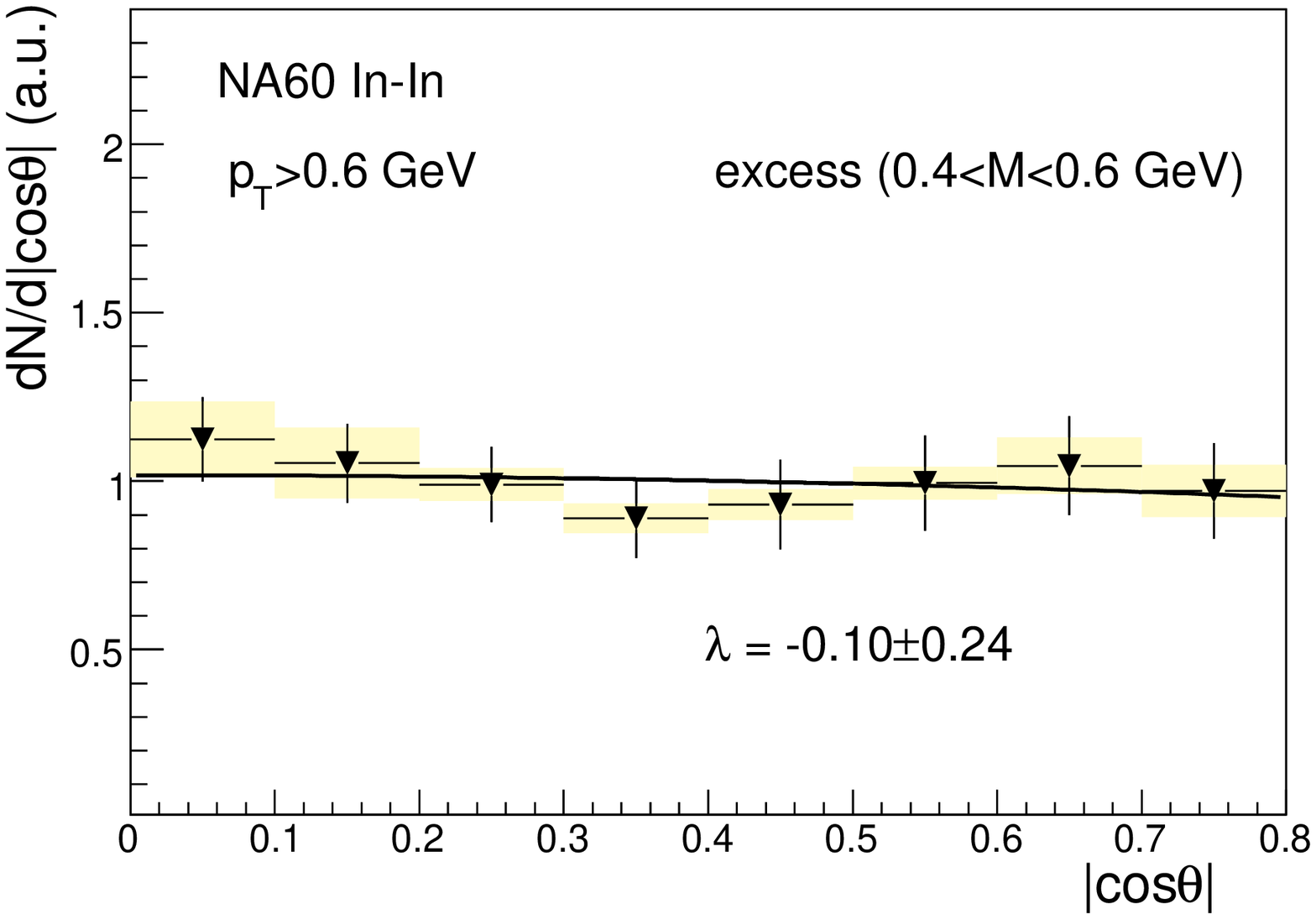}	 
\includegraphics[scale=0.32]{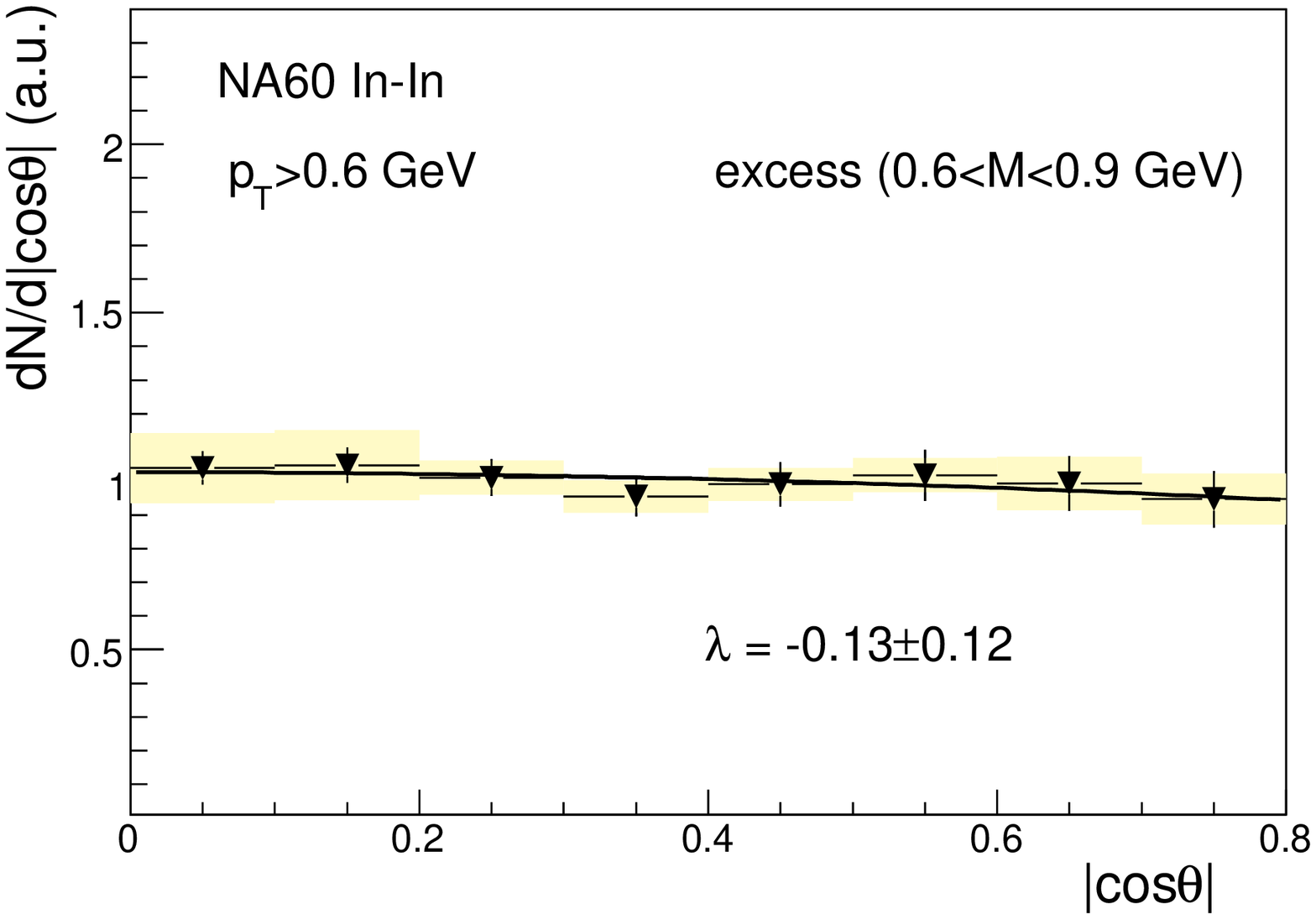}
\caption[]{Polar angle distributions for excess dileptons.}
\label{Fig2}
\vspace{-0.6cm}
\end{figure}
\begin{figure}[ht]
\centering
\includegraphics[scale=0.32]{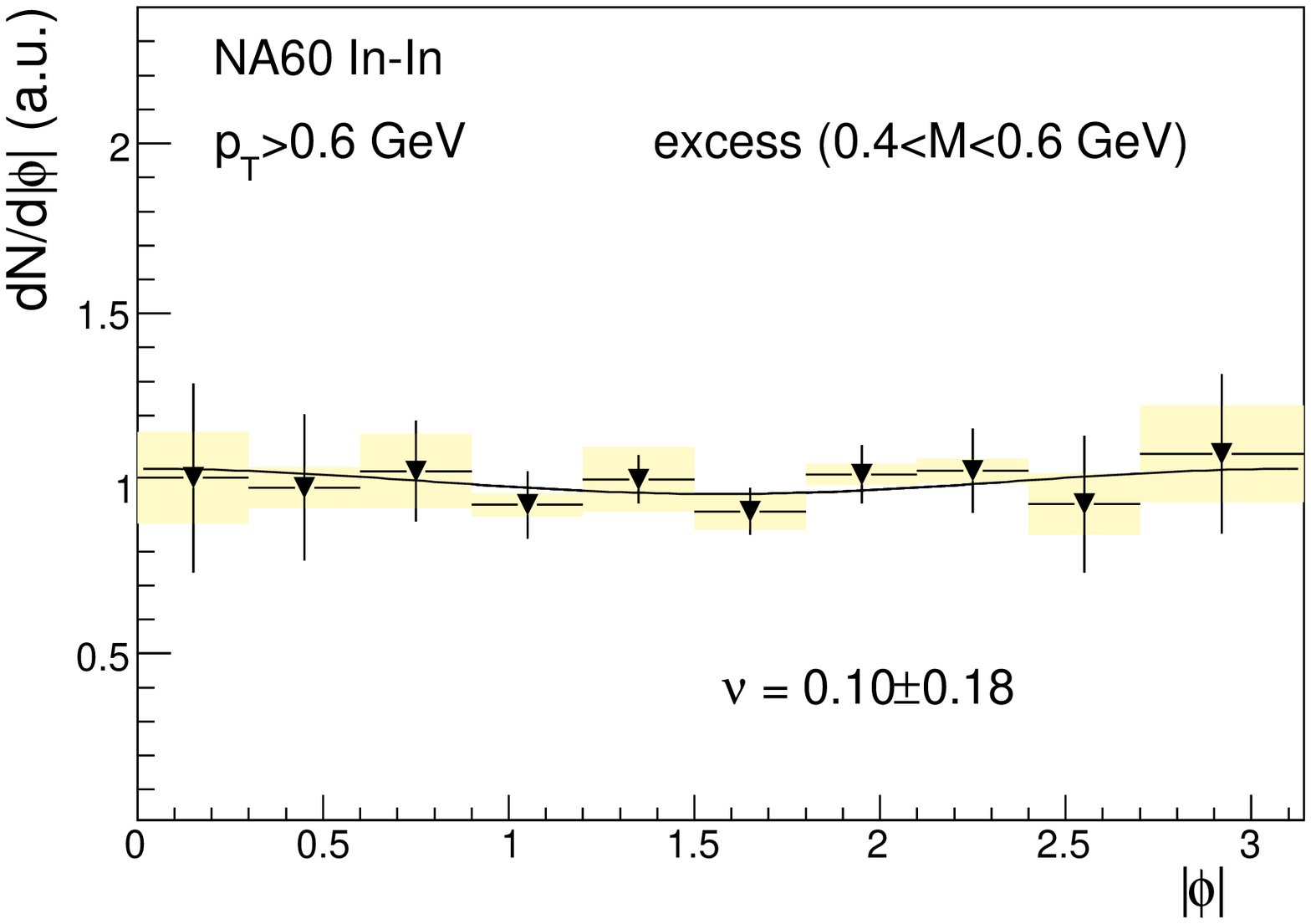}	 
\includegraphics[scale=0.32]{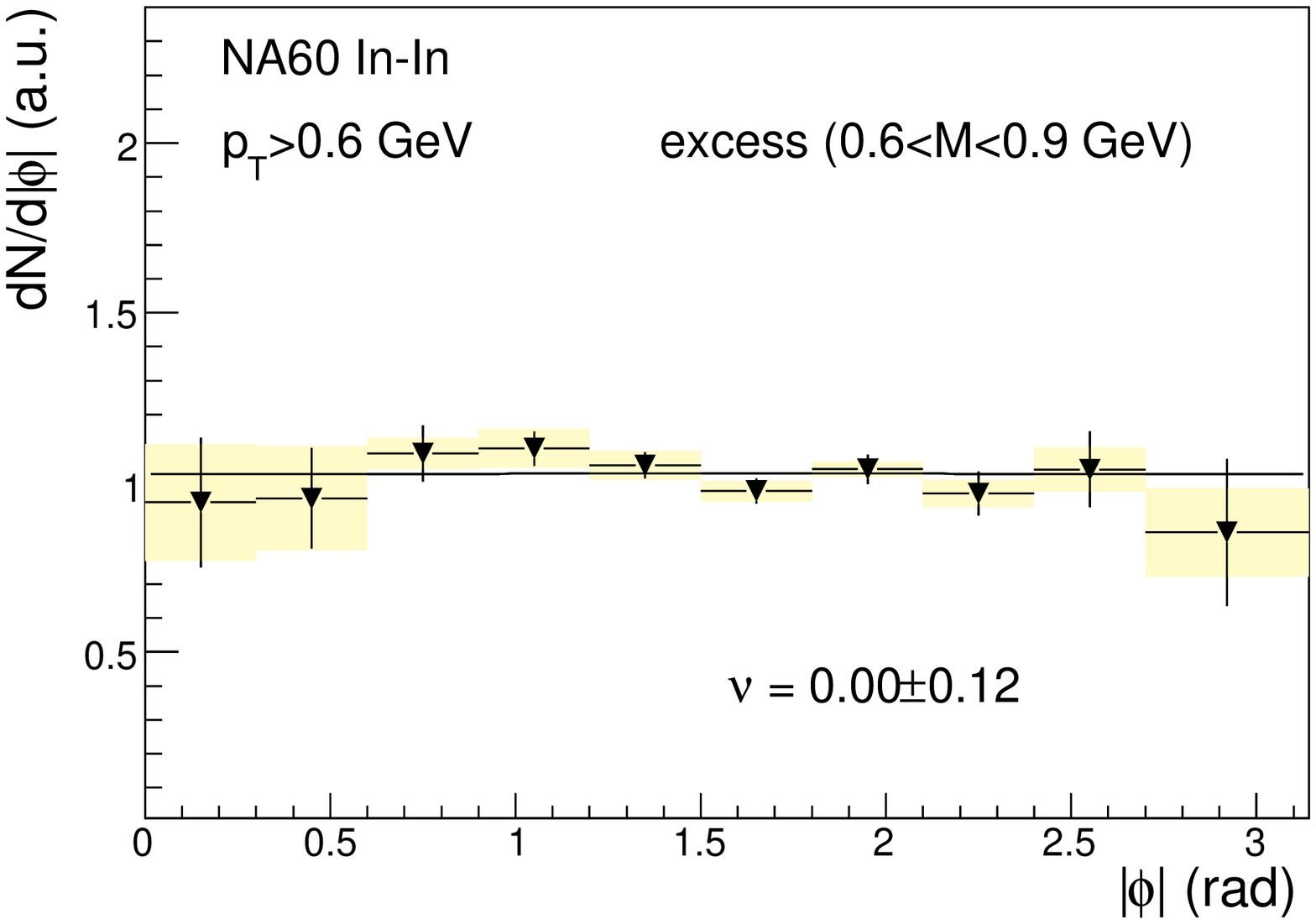}
\caption[]{Azimuth angle distributions for excess dileptons.}
\label{Fig3}
\end{figure}

The left panel of Fig.~\ref{Fig1} shows the 2-dimensional contour
plot from the fit for  the excess dileptons in the two mass
intervals $0.6<M<0.9$ GeV, while the right panel shows the
corresponding projections in $\cos\theta$ for the different bins in
$\phi$. Similar results are obtained for the mass interval $0.4<M<0.6$
GeV. The parameter $\mu$, which can be determined only from these
bidimensional fits, is consistent with zero in both cases: $\mu = 0.05
\pm 0.03$ for $0.6<M<0.9$ GeV and $\mu = -0.04 \pm 0.10$ for
$0.4<M<0.6$ GeV.
%
%
The parameters $\lambda$ and $\nu$ are determined with a better
statistical accuracy from the analysis of the 1-dimensional
projections where the low populated bins in the acceptance borders
have a smaller effect.  The polar and azimuth distributions obtained
from methods 2 and 3 are shown in Fig.~\ref{Fig2} and Fig.~\ref{Fig3},
respectively.  Also $\lambda$ and $\nu$ are consistent with zero for
both mass windows.
%
\begin{figure}[ht]
\centering
\includegraphics[scale=0.3]{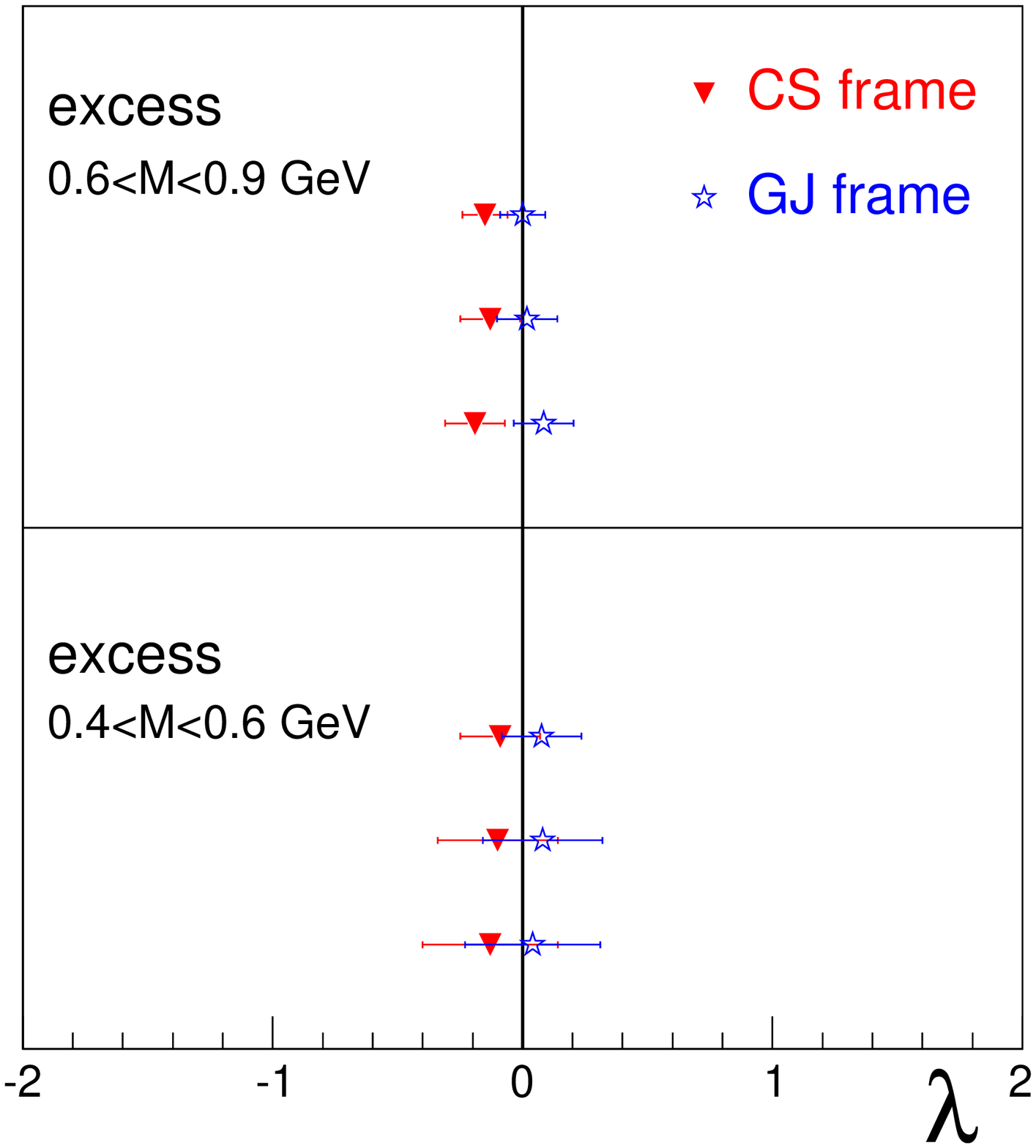}
\includegraphics[scale=0.3]{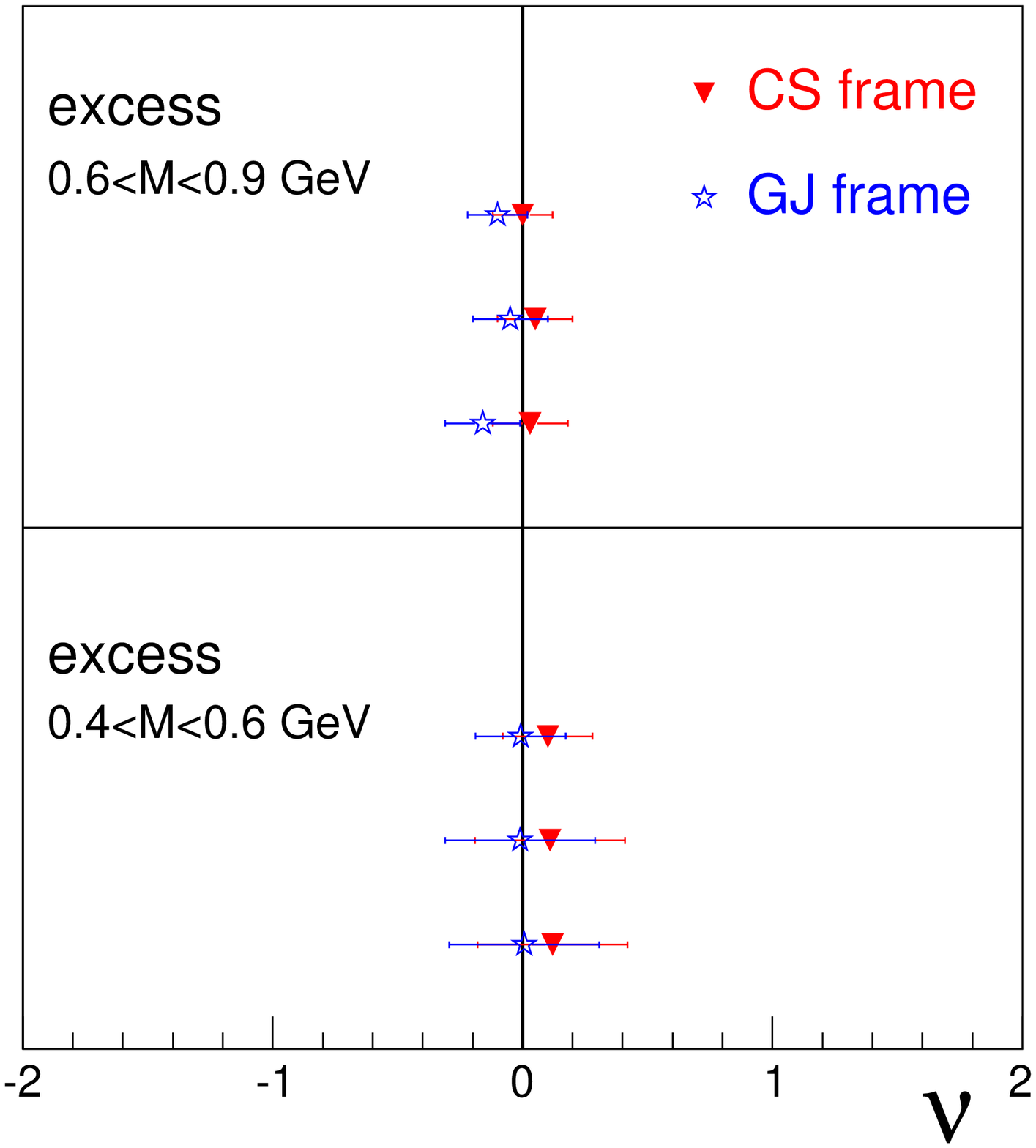}
\caption[]{Structure functions $\lambda$ and $\nu$ in the
Collins-Soper and Gottfried-Jackson frames. }
\label{Fig4}
\end{figure}

As already mentioned, once all measured, $\lambda$, $\mu$ and $\nu$
can be re-computed in any other frame. Fig.~\ref{Fig4} shows the
values of $\lambda$ and $\nu$ determined from the fits in the CS frame
and in the Gottfried-Jackson frame as obtained by applying the
transformation of ~\cite{Falciano}. The structure functions remain
consistent with zero independent of the frame.

The systematic error due to the combinatorial background subtraction
is $\sim2-3\%$. The systematic error due the fake matches is less than
1\%. Variations in $M$, $y$ and $p_T$ used as input for the Monte
Carlo lead to uncertainties much smaller than the statistical errors.
Systematic uncertainties in the isolation of the excess range from
$4-6\%$ up to $10-15\%$ in some low-populated $\cos\theta-\phi$ bins
and they represent the main source of systematic errors.  However, the
measurement still remains dominated by the statistical errors.  If the
systematic errors are conservatively assumed uncorrelated from point
to point and added in quadrature to the statistical fit errors quoted
in the numbers above, this would increase by 15-20\%.

Summarising, the absence of any polarization is fully consistent with
the interpretation of the observed excess as thermal radiation. It is
important to remember that this is a necessary but not sufficient
condition.  Put together with the other features - Planck-like shape
of mass spectra, temperature systematics, agreement of data with
thermal models - this makes the thermal interpretation more plausible
than ever before.


%
%
%
%
\end{document}